\documentstyle[titlepage,12pt]{article}

\title{\bf Numerical Studies of the Two Dimensional XY Model
        with Symmetry Breaking Fields}

\author{ T. Ala--Nissila$^{1,4,5}$, E. Granato$^2$,
K. Kankaala$^{3,5}$,\\
J. M. Kosterlitz$^{4}$, and  S.--C. Ying$^{4}$ \\ \\
$^1$Research Institute for Theoretical Physics \\
University of Helsinki \\ P.O. Box 9 (Siltavuorenpenger 20 C) \\
FIN--00014 University of Helsinki, Finland \\ \\
$^2$Laborat\'orio Associado de Sensores e Materiais \\
Instituto Nacional de Pesquisas Espaciais \\
12225 - S\~ao Jos\'e dos Campos, S\~ao Paulo, Brazil \\ \\
$^3$Center for Scientific Computing \\
P.O. Box 405, FIN--02101 Espoo, Finland \\ \\
$^4$Department of Physics, Brown University \\
Box 1843, Providence, R.I. 02912, U.S.A. \\ \\
$^5$Department of Electrical Engineering \\
Tampere University of Technology \\
P.O. Box 692, FIN--33101 Tampere, Finland \\
}

\date{July 13, 1994}

\begin{document}

\maketitle

\begin{abstract}
We present results of numerical studies of the
two dimensional $XY$ model with four and eight fold
symmetry breaking fields. This model has recently been shown
to describe hydrogen induced reconstruction on the $W(100)$ surface.
Based on mean--field and renormalization group arguments,
we first show how the interplay between the
anisotropy fields can give rise to different phase transitions in
the model. When the fields are compatible with each other
there is a continuous phase transition when the fourth order
field is varied from negative to positive values.
This transition becomes discontinuous at low temperatures.
These two regimes are separated by a multicritical point. In the case of
competing four and eight fold fields, the first order transition
at low temperatures opens up into two Ising transitions.
We then use numerical methods to accurately locate the position
of the multicritical point, and to verify the nature of the transitions.
The different techniques used include Monte Carlo histogram
methods combined with finite size scaling analysis,
the real space Monte Carlo Renormalization Group method,
and the Monte Carlo Transfer Matrix method. Our numerical
results are in good agreement with the theoretical arguments. \\

\end{abstract}

\textheight 21cm
\textwidth 14.5cm
\oddsidemargin 0.96cm
\evensidemargin 0.96cm
\topmargin -0.31cm
\raggedbottom
\baselineskip 24pt
\parindent=0mm

\section{Introduction}

In two dimensions, conventional long range order
cannot exist in continuous spin
models ($O(n), n\geq 2$) because it is destroyed by spin wave excitations
\cite{Kos73,Mer66,Mer68}. However, Kosterlitz and Thouless
\cite{Kos73} proposed that in the $XY$ ($O(2)$) model
there is a phase below the critical temperature
where topological long range order can be defined. The vanishing of
this order occurs via the Kosterlitz--Thouless (KT) transition.
Physical systems where the KT transition occurs are
numerous; they include
superfluid $^4${\it He} films, Josephson junction arrays,
superconducting transition of type II, and various
phase transitions on surfaces and adsorption layers.
\medskip

In many cases, the realization of the $XY$ model is accompanied by
various symmetry breaking fields, whose effect is very
complicated as demonstrated qualitatively by Jos\`e {\it et al.}
\cite{Jos77}. For example, it was recognized already at an
early stage \cite{Jos77} that the presence of a four fold
field restores a conventional phase transition, but with
continuously varying critical exponents. In contrast, a
six fold symmetry breaking field opens up the KT transition into
two parts: at high temperatures, the transition remains $XY$
type, but at low enough temperatures it is into a discrete
planar phase in which the system orders along one of the
six preferred directions \cite{Sel88}.
Both of these situations have
been realised in experimental systems; four fold fields
are know to be present in surface structural phase
transitions whereas liquid crystals provide a case where
the six fold field exists.
\medskip

The effects of symmetry breaking fields are further complicated
by the existence of {\it higher order} multiples of the
fields, which are allowed by symmetry. In most cases, since
these fields are irrelevant if lower order fields are present,
their influence has been neglected. However, if it happens
that the lowest  order symmetry breaking field vanishes,
the higher harmonics can become
relevant at low enough temperatures \cite{Jos77}. A demonstration
of this fact is the study of Selinger and Nelson \cite{Sel88}
who modeled a phase transition occuring in liquid crystals
by an $XY$ model with six and twelve fold symmetry breaking
fields. They found a rich behavior of the phase diagram
depending on whether or not the two fields are compatible
with each other.
\medskip

We have recently developed \cite{Kan93b} a lattice Hamiltonian
for the adsorption system $H/W(100)$, which is based on
the $XY$ model with a four fold symmetry breaking field,
and its higher harmonics. We have argued that
the essential physics of this model is dictated by an
interplay between the four fold and the eight fold fields,
in a manner very similar to that of Ref. \cite{Sel88}.
The intriguing aspect of this model is that the strengths
of the symmetry fields --- the four fold field in particular ---
are {\em tunable} by changing the amount of adsorbed hydrogen.
In fact, it was demonstrated that the fourth order field {\it
vanishes} at a hydrogen coverage of about 0.1 for this
system. This system thus provides an ideal example
for studying the effect of interplay between symmetry
breaking fields within the $XY$ model.
\medskip

The purpose of the present work is to conduct a
detailed, quantitative study of the two dimensional $XY$
model with four and eight fold anisotropy fields.
We shall first discuss in detail, how the interplay between these
two anisotropy fields dictates the nature of the phase
diagram at low temperatures where the eight fold
field is relevant in the
renormalization group sense. We
give both mean field and renormalization
group arguments in explaining how it is possible to obtain either
a discontinuous, or two continuous Ising transitions
at low temperatures due to the interplay between the anisotropy fields.
Namely, when
the four and eight fold fields are compatible with each other and do not
compete, this gives rise to a first order transition at $h_4=0$ as we pass
from negative (positive) values of the four
fold field to positive (negative)
with a finite $h_8$ field. However, when the two
fields are competing with each
other, the first order transition opens up into two Ising transitions
with an intermediate phase in between.
\medskip

Following analytic arguments,
we proceed to simulate the $XY$ model with symmetry breaking fields
using the Monte Carlo method with the Wolff updating algorithm
 which is generalized to
include contributions from the anisotropy fields.
We employ finite size scaling arguments
to locate the multicritical point in this model.
These results are further
corroborated by Monte Carlo Transfer Matrix (MCTM) studies.
We also verify the existence
of the continuous Ising transition
in the case of competing anisotropy fields by
using both the MCTM and real space Monte Carlo Renormalization Group
methods. Our results are in good agreement with theoretical predictions,
and also consistent
with available experimental data for the $H/W(100)$
adsorption system.

\section{Model Hamiltonian and Qualitative Renormalization Group Analysis}

The Hamiltonian of the $XY$ model with four and
eight fold anisotropy fields can be written as

\begin{equation}
\label{eq:xyh4h8}
H = -K\sum_{<i,j>} \cos (\phi_i - \phi_j) - h_4\sum_i\cos (4\phi_i )
                                          + h_8\sum_i\cos (8\phi_i ),
\end{equation}

where $K$ is the $XY$ coupling constant between
nearest neighbors, $\phi_i$ are the angle variables defined by the
individual spin vectors $\vec \sigma_i = (\cos\phi_i,
\sin\phi_i)$ on site $i$,  $h_4$ and $h_8$ are the
four and eight fold symmetry breaking fields,
respectively, and we have subsumed
the temperature into the coupling constants and fields.
The summation $<i,j>$ goes over the nearest
neighbors, and the summations $i$ are over all lattice sites.
\medskip

We will first discuss mean field theory
to obtain a qualitative picture of
the critical phenomena that Eq. (\ref{eq:xyh4h8})
gives rise to. Our purpose is to give insight to the
underlying physics which is dictated by the
interplay between the anisotropy
fields. These arguments correspond to the case where both fields
are always assumed to be relevant, and will thus give correct
qualitative behavior at low temperatures only.
There are two possible scenarios depending on the anisotropy
potential $V(\phi)$ of Eq. (\ref{eq:xyh4h8}), defined by

\begin{equation}
\label{eq:potential}
V(\phi ) \equiv - h_4\cos (4\phi) + h_8\cos (8\phi).
\end{equation}

Namely, when
$h_4$ is negative it favors spins aligning along the directions
$\phi=\pi/4 +n\pi/2, n=0,1,...$ whereas a positive $h_4$ favors
$\phi = n\pi/2, n=0,1,...$\ .
The eight fold field has two possibilities as well. When $h_8 <0$ the
favored orientations are $\phi=n\pi/4, n=0,1,...$ which are exactly
the directions favored by either a negative or a positive four fold field.
We say that in this case the anisotropy fields
are {\em non--competing} or {\em compatible}
with each other. For $h_8 >0$ the favored orientations are
$\phi=\pi/8 +n\pi/4, n=0,1,...$\ . As these are
different from those favored by the four fold
field, {\em competition } will set in when the four and
eight fold fields are of the same order of magnitude.
\medskip

We first consider the case of a negative, {\em i.e.} non--competing
$h_8$. In Fig. 1(a),
we show the anisotropy potential $V(\phi )$ as a
function of the angle $\phi $ for various values of $h_4$,
given a fixed $h_8$.  The local minima at $\phi = 0$,
and at $\phi = \pm \pi/4$ are clearly visible. When $h_4 > 0$,
the minima at $\phi = 0$ are deeper, whereas for
$h_4 < 0$ the minima at  $\phi = \pi/4$ are deeper. As $h_4$ passes
through zero from $h_4 > 0$ to $h_4 < 0$, there is a first order
transition from one minimum to another.
\medskip

In the case of a positive or competing eight fold field,
the situation is more complicated.
The behavior of the potential $V(\phi )$ for fixed $h_8>0$ and various
values of $h_4$  is shown in Fig. 1(b).
For $h_4 > 4\vert h_8\vert $, there is only one minimum
at $\phi = 0$, and one at $\phi = \pm \pi/4 $
for $h_4 < -4\vert h_8\vert $.
But now as the four fold field
passes through $h_4 = 4\vert h_8\vert$, the single minimum splits into
two at $\phi = \pm \phi^{\prime}_0$. By minimizing the anisotropy
potential $V(\phi)$,
we can show that these new minima begin to form at

\begin{eqnarray}
\phi & = & \pm \frac{1}{4}\vert \cos^{-1}\left
(\frac{h_4}{4h_8} \right ) \vert
                       \nonumber \\
     & \equiv  & \pm \phi^{\prime}_0.
\end{eqnarray}

At the other boundary where $h_4$ passes through $-4\vert h_8\vert$,
the same argument applies except that $\phi^{\prime}_0$ now measures
the deviation from $\pi/4$.
The first order transition for the compatible field case
(cf. Fig. 2(a)) has now opened up into two continuous transitions with an
intermediate phase in between (cf. Fig. 2(b)).
\medskip

We will next show explicitly that the
continuous phase transition belongs
to the universality class of the two dimensional Ising model.
We can approach the transition boundaries
either from the phase where $h_4>0$
where the preferred orientation
of spins is along  $\phi = 0,\pi /2,...$ \ ,
or from the phase where $h_4<0$ and the favored directions are then
$\phi=\pi /4,3\pi /2,... $\ . Let us consider the former possibility.
For large $h_4$, the system is in the energy minimum
at $\phi=0$. When we approach the transition ($h_4 \rightarrow 0$),
two minima form at $\phi = \pm \phi^{\prime}_0$.
Let us consider the contribution to the partition
function due to these new minima.
The Boltzmann weights can be written as
$ \exp \{ K\cos[(s_i - s_j)\phi_0^\prime]\}$ where $s_i = \pm 1$.
We denote these weights by $W[s]$ where $s = s_i - s_j = 0,2$.
For two neighboring spins at the same minimum,

\begin{equation}
 W[0] = \exp (K),
\end{equation}

while for two opposite spins,

\begin{equation}
 W[2] = \exp (K\cos 2\phi_0^\prime).
\end{equation}

The equivalent Ising coupling constant is thus
$J = K\sin^2\phi_0^\prime$. Since the argument is valid in the limit
$K\rightarrow \infty$ only, we will later numerically verify the
nature of the Ising transition predicted here.
\medskip

Next, we present more quantitative
renormalization group (RG) arguments
which are a direct extension of the work by Selinger and
Nelson \cite{Sel88}.
At high temperatures (where $K$ is small),
only the four fold field is relevant
in the RG sense, and the long range order in the system is dictated
by $h_4$ alone. At some temperature corresponding to $K_m$ the
eight fold field becomes relevant. For lower temperatures
(larger values of $K$)
and for finite $h_4$ and $h_8$, the nature of the phase
diagram is determined by the interplay between these two fields
as qualitatively discussed above.
Any anisotropy field $h_p$ of order $p$
will obey the RG recursion relation \cite{Jos77}:

\begin{equation}
\label{eq:rgh}
 h^{'}_p = b^{\lambda_p}h_p,
\end{equation}

where $h^{'}_p$ is the field obtained after an RG iteration, $b$ is a
constant, and

\begin{equation}
\label{eq:relevantt}
 \lambda_p = 2 - \frac{1}{4\pi K} p^2.
\end{equation}

When the parameter $\lambda_p > 0$,
the field $h^{'}_p$ will increase as iterations proceed, and
the  $p^{th}$ anisotropy field is a {\em relevant}
variable \cite{Jos77}. However, as the temperature
increases, higher order fields become {\em irrelevant}
($\lambda_p < 0$),
and the respective  $h^{'}_p$ will decrease with iteration.
On the other hand, it can easily be seen that at $K=\infty$
all symmetry breaking perturbations are relevant.
If $\lambda_p = 2 - p^2/(4\pi K) = 0 $, we get the
temperature $K^{-1}_{m,p}$ below which the field
$h_p$ is relevant, {\em i.e.}
it is {\em marginally} relevant at that point \cite{Jos77}:

\begin{equation}
\label{eq:rglambda}
   K_{m,p}^{-1} = \frac{8\pi}{p^2}.
\end{equation}

We can conclude that the four and eight fold anisotropy fields
are relevant at all
temperatures $K^{-1} < K_{m,4}^{-1} = \pi/2$ and $K^{-1}
< K_{m,8}^{-1} = \pi/8$, respectively.
The results for the $XY$ model with only the four fold
field $h_4$ are well known \cite{Jos77}.
The continuous phase transition into
an ordered state ($h_4 \ne 0$)
along the critical lines terminating
at $K^{-1}_{KT} < K_{m,4}^{-1}=\pi/2$
belongs to the universality class of the $XY$ model with a cubic
anisotropy field, where
$K_{KT}^{-1}$ is the (true)
order--disorder transition temperature for the
pure $XY$ model. Along the critical line $h_4=0$ we have a series of
continuous $XY$ transitions for all $K^{-1} < K_{KT}^{-1}$
(cf. Fig. 2). In the other limit where
$h_4 \rightarrow \infty$, the model becomes a four state
clock model which can be shown to decouple into two Ising models
\cite{Bet64}.
\medskip

With the inclusion of the eight fold field, different
scenarios occur depending on the signs of $h_4$ and $h_8$.
For the remainder of this paper, we will denote
by $K^{-1}_m$ the true value of the coupling constant,
where the eight fold field becomes relevant.
For $K^{-1} < K^{-1}_m$, both $h_4$ and $h_8$ are relevant.
If $h_8$ is positive, the renormalized potential $V^*(\phi)$ is similar to
that of Fig. 1(b) with two continuous
Ising transitions. In the limit $K \rightarrow \infty$,
we obtain the exact mean field result: the transition
occurs at $h_4 = \pm 4\vert h_8\vert$.
In the other limit $K \rightarrow K_m$,
it can be shown \cite{Sel88} that to
lowest order in $\vert h_8 \vert$, we have

\begin{equation}
K_m^{-1} = \frac{\pi}{8} + B\vert h_8\vert ,
\end{equation}

where $B>0$ is a constant which can be estimated to be ${\cal O}(1)$.
The upper bound for $K^{-1}_m$ is the
Kosterlitz--Thouless transition temperature $K^{-1}_{KT}<\pi/2$
for this model.
The lower bound, on the other hand, is given by the zero field estimate
$K_m^{-1}(h_8=0) = \pi /8$.
The corresponding phase diagram is shown schematically in Fig. 2(b).
\medskip

For a negative $h_8$ and for $K^{-1} < K^{-1}_m$, the system fluctuates in
the minima of Fig. 1(a). The location of the deepest
minimum changes as $h_4$ passes through $h_4 = 0$, and we expect a
first order phase transition.
Selinger and Nelson \cite{Sel88} have in fact shown that
in this case there is a discontinuity in the
order parameter across the transition. This discontinuity
vanishes exponentially as $K \rightarrow K_m$ \cite{Sel88}
thus making it very difficult to numerically locate $K_m$.
The phase diagram corresponding to this non--competing case is
depicted in Fig. 2(a).
\medskip

In the remainder of this paper, we will perform detailed
numerical studies of two particular aspects of the phase
diagrams shown in Figs. 2(a) and 2(b). The first concerns
the exact location of the multicritical point $K_m$ for
$h_4=0$, for a finite $h_8$. The second is the verification
of the Ising--like nature of the low temperature transition
lines in Fig. 2(b) for the competing case.

\section{Location of the Multicritical Point}

To quantitatively locate the multicritical point $K_m$, we
have performed extensive Monte Carlo simulations of Eq. (1),
by adapting a modified Wolff algorithm.
The Wolff algorithm was originally developed
for isotropic, continuous
spin systems such as the $XY$ model \cite{Wol89a}. In our case, we
have added symmetry breaking fields to the model.  This is accounted
for by modifying the Wolff cluster update algorithm in the following way.
We divide the Hamiltonian of Eq. (\ref{eq:xyh4h8}) into two parts, the
isotropic part $H_{XY}$,

\begin{equation}
H_{XY} = -K\sum_{<i,j>} \cos (\phi_i - \phi_j),
\end{equation}

and the anisotropic part $H_{4,8}$,

\begin{equation}
H_{4,8} = - h_4\sum_i\cos (4\phi_i ) + h_8\sum_i\cos (8\phi_i ).
\end{equation}

We form the Wolff cluster for the isotropic part $H_{XY}$ in the usual
fashion using a cluster labeling technique
similar to the ``ants in the labyrinth'' scheme \cite{Bai91b}. We then
calculate the change in the energy due to the anisotropy fields for
the old and the proposed new cluster as

\begin{equation}
\Delta H_{4,8}=H_{4,8}^{new} - H_{4,8}^{old}.
\end{equation}

Whether or not the cluster is flipped is determined by applying the
standard Metropolis acceptance criterion to this energy difference.
It is easy to verify that this combined algorithm satisfies detailed
balance and is ergodic.
\medskip

It has recently been shown
\cite{Kan93thesis} that the Wolff algorithm
performs poorly for anisotropic $XY$
models at low temperatures. It probes the
phase space effectively but in a
presence of strong anisotropy fields, reaching
thermal equilibrium from an initial state can take very long. Metropolis
algorithm, on the other hand,
reaches local equilibrium rapidly but fails to
search the phase space extensively.
We have overcome these problems by
a scheme where Wolff and Metropolis algorithms are
simply combined by inserting
several Metropolis local update sweeps after
a certain amount of Wolff steps. Similar approach
has also been suggested previously in Ref. \cite{BhaXX}.
\medskip

To locate the multicritical point of the $XY$ model with anisotropy
fields, we use the method of Lee and Kosterlitz \cite{Lee90}.
In this scheme, the ``free energy'' $F$ for a given system of
linear size $L$ with order parameter $\Psi$ and with periodic boundary
conditions can be expressed as

\begin{equation}
\label{eq:defF}
\exp [-F(\Psi ,L, N)] = N Z^{-1}(\beta)
\sum_E \Omega (E,\Psi )\exp (-\beta E),
\end{equation}

where $N$ is the number of samples (configurations), $Z(\beta)$
is the partition function,
$\Omega (E,\Psi )$ is the number of states with energy $E$, and
we assume that the transition is driven by an external field.
In the following, we shall for simplicity drop the $N$ dependence
of $F$.
$F$ differs from the actual bulk free energy
but its {\em shape} is identical to that of
the bulk free energy and thus is also the {\em difference} $\Delta F$
(see Eq. (\ref{eq:DeltaF}) below).
The ``free energy'' has two minima due to the two coexisting phases.
These minima, located at $\Psi_1$ and $\Psi_2$, are separated by a
maximum at $\Psi_m$. In the thermodynamic limit the
double minima structure vanishes at the transition points but in a finite
system it may persist even above the transition temperature. In this
method it is precisely this property that is exploited to reveal the
order of the transition for a finite system.  More specifically, the
free energy can be expanded below the transition where the correlation
length $\xi \ll L$ as

\begin{equation}
\label{eq:expF}
F(\Psi,L) = L^df_0(\Psi ,g) + L^{d-1}f_1(\Psi ,g) + \ldots ,
\end{equation}

where $f_0(\Psi ,g)$ is the bulk free energy density, $f_1(\Psi
,g)$ is a surface term which has a maximum at $\Psi_1 < \Psi_m <
\Psi_2$, and $g$ is a scaling field $g \propto (T-T_C)$. It can
then be shown that the free energy has a minimum on both sides of the
maximum $\Psi_m$, and that the height difference between the minima and the
maximum is

\begin{equation}
\label{eq:DeltaF}
\Delta F(L) \equiv F(\Psi_m,L) - F(\Psi_1,L) = A(g)L^{d-1} +
                                                   B(g)L^{d-2} +
\ldots  \ .
\end{equation}

This expansion holds for first order transitions when $\xi \ll L$, and
the free energy difference $\Delta F(L)$ is an increasing function of
the system size $L$. Even for $L
\ll \xi$, $\Delta F(L)$ is an increasing function of $L$. Thus, at the
first order transition point when
$F(\Psi_1,L) = F(\Psi_2,L)$, $\Delta F(L)$
is an increasing function of $L$. In the
disordered phase, the free energy difference $\Delta F(L)$ {\em
decreases} as a function of $L$.

\medskip
Near a fixed point describing a continuous transition, a scaling
form can be developed for the singular part of the free energy
\cite{Lee90}. Its analytic expansion gives $\Delta F(L) =
a - bgL^{1/\nu} + {\cal O}(g^2L^{2/\nu})$, where $a$ and $b$
are $L$--independent constants. This form is appropriate for
$L \ll \xi$, where $g>0$ for $K<K_m$ and $g<0$ for $K>K_m$.
Thus, we expect that $\Delta F(L)$ increases with $L$ in
the low temperature phase. However, for our model the behavior
in the vicinity of $K_m$ is expected to be very complicated,
and a finite size scaling form has not been developed. Naively,
we would expect that because $1/\nu=0$ and $\ln(\xi) \sim g^{-1/2}$,
the barrier $\Delta F(L)$ will increase as $g \ln^2(L)$. For
$K>K_m$, however, $\Delta F(L)$ must increase more
rapidly with $L$ eventually
crossing over to the linear behavior of Eq. (15) for $L \gg \xi$
deep into the first order regime.
We will use this property of $\Delta F(L)$ and change $K$
at $h_4=0$ to locate $K_m$, as explained below.

\medskip
To facilitate the use the finite size scaling technique of Lee and
Kosterlitz, we calculated the histogram of the order parameter
$\Phi=(1/L^2)\sum_i \cos 4\phi_i$ summed over the lattice
by dividing the interval $[0,1]$ into 200 equal
bins, and putting each value into its respective bin. Thus,
we can construct a histogram which shows the
double peak structure. This
histogram is essentially an approximation of the partition function.
By taking the negative of the logarithm of this histogram, we obtain
an approximation for the free energy distribution $F(\Psi ,L)$ of the
system (cf.  Eq. (\ref{eq:defF})). As we reside on the transition line
at $h_4=0$, the two peaks of $F(\Psi ,L)$ are equally high, and we
can readily calculate the difference $\Delta F(L)$ given in Eq.
(\ref{eq:DeltaF}). For the eight fold field, we used the value $h_8 =
-0.15$.
When using the combined algorithm, we first did 3000 Metropolis
steps to reach a local equilibrium and then continued with 5000 Wolff
cluster formations with 10 Metropolis steps after each 1000 cluster
formations. All this information was discarded. The data were
averaged over 1 000 000 cluster formations so that after every 1000 Wolff
cluster formations ten Metropolis steps followed.  We
calculated the order parameter histogram for several systems of sizes
$L=8$, $12$, $16$, $24$, $32$, $48,$ and $64$ at various temperatures
along the line $h_4=0$. From
these data we can then deduce $\Delta F(L)$.
In the first order regime and in two dimensions,
it should scale linearly
with increasing $L$. If the transition is continuous, the double peak
structure should vanish in the limit $L \rightarrow \infty$.  We also
studied the distribution of angles by constructing a histogram of each
individual angle $\phi_i$.
\medskip

Our main results are depicted in Figs. 3(a)--(c).
Typical histograms for various systems sizes are shown in
Figs. 3(a) and (b), and the extracted energy barriers
$\Delta F(L)$ as a function of the linear system size $L$ in Fig. 3(c),
for the present case of non--competing anisotropy
fields. All the $\Delta F(L)$'s were calculated by fitting an eight
order polynomial to the data.
All data points are averages of about $10^6$
configurations. The energy difference increases with
$L$ for temperatures corresponding to
$K \geq 2.3$, which indicates a first order
regime. The behavior of $\Delta F(L)$ at
temperatures corresponding to
$K \leq 2.2$ indicates, on the other hand, a
regime where the transition is continuous.
At $K = 2.2$, $\Delta F(L)$ first seems to increase
with $L$ up to $L=12$ and then
decrease for larger $L$, although within error bars it's
almost constant. At $K=2.1$, no double
peak structure exist. We also analysed the size dependence of
the multicritical point $K_m$. By fitting
$\Delta F(L)$ {\em vs.} the logarithm of
inverse temperature $K$ for $L=8,12,16,24,48,64$, we were able
to determine the multicritical
point $K_m(L)$ for each system size. From these we estimated the
multicritical point by scaling this data against $1/L$.
We conclude that

\begin{equation}
\label{eq:multiMC}
K_m = 2.1 \pm 0.1,
\end{equation}

which is the main result of this section. It is also well within the
theoretical bounds $2/\pi < K_m < 8/\pi$, as expected.
\medskip

Finally, we should note that
the accuracy of the result suffers from severe fluctuations in the
vicinity of the multicritical point, in particular
for the largest systems. This inhibits the use of
histogram techniques for extrapolation \cite{Fer88}.

\section{Ising Transition in the Case of Competing Anisotropy Fields}

The other scenario for our model is the case where the anisotropy
fields are competing. It was shown analytically that the
first order transition at low temperatures opens up to two Ising
transitions with an intermediate phase in between. In this intermediate
phase the long range order is dictated by the eight fold field.
To see this transition numerically, we chose a finite
four fold field $h_4=0.06$ which favors the orientations
$\phi_i = 0, \pi/2$, etc. for the individual spins.
The competing
eight fold field is chosen to be equal in magnitude
to the one used in the
non--competing case {\em i.e.} $h_8 = 0.15$. We try to locate the
corresponding Ising
transition temperature $K_I^{-1}$ by scanning the inverse temperature $K$.
$K_I$ should be well above our estimate of
$K_m\approx 2.1$
(cf. Fig. 2(b)). For this calculation, we used
the Monte Carlo renormalization group
(MCRG) scheme proposed by Binder \cite{Bin81}.
Consider the $XY$ spins $\vec
\sigma_i=(\cos\phi_i,\sin\phi_i)$ on a
two--dimensional lattice which
is divided into subcells or blocks of (linear) size $L_B$. Let us first
define a block variable

\begin{equation}
\Phi_{L_B} = \frac{1}{L_{B}^2}\sum_{i \in L_B} \psi_i,
\end{equation}

where $\psi_i$ is a measure for the local order in the system.  We can
then define an order parameter for each block size as $\Psi_{L_B} =
<\Phi_{L_B}>$ where brackets again denote a configurational average.
We studied different moments of the block variables $\Phi_{L_B}$, and
constructed the fourth and sixth order cumulants $U_{L_B}$ and
$V_{L_B}$ for each block size as in Ref. \cite{Bin81}.
The variation of these two cumulants as a function of the block size
$L_B$ gives a flow diagram analogous to that of a renormalization
group method. These cumulants
approach zero above $T_C$ as the block size increases. Below $T_C$,
both cumulants tend to nonzero values $U_{L_B} \rightarrow {2/3}$, and
$ V_{L_B} \rightarrow {8/15}$ as $L_B \rightarrow \infty$.  At the
critical point $T_C$, the cumulants approach nontrivial fixed point
values $U^*$ and $V^*$. Thus, the behavior of the
cumulants is reminiscent of the renormalization group flows under
subsequent transformations of the length scale.
One can also estimate the correlation length exponent $\nu$
from the data in the vicinity of $U^*$ by noting that \cite{Bin81}

\begin{equation}
\label{eq:exponent}
\frac{U_{L^\prime_B} - U^*}{U_{L_B}-U^*} \simeq b^{(1-\alpha)/\nu},
\end{equation}

for subsystem blocks of size $L_B^{\prime}=L_B/b$,
and by using the scaling relation $\alpha = 2 - \nu d$.
\medskip

For the block variable we chose

\begin{equation}
\Phi_{L_B} = \frac{1}{L_B^2}\sum_i \left (\sin \phi_i \right ).
\label{eq:sincos}
\end{equation}

This order parameter, when the angles
are folded between $-\pi/4 < \phi < \pi/4 $, is
zero in the high temperature phase and finite in the low temperature phase
when a single domain dominates below and above the transition.
We studied the fourth and sixth order cumulants $U_{L_B}$ and
$V_{L_B}$. From the flows of these cumulants as a function of the
inverse linear size of the block, we can deduce the nontrivial fixed point
value from which we can further extract $K_{I}$.
\medskip

At low temperatures, the simulations suffer from  high barriers
between different regions of the phase space, which we tried to
overcome by using the combined algorithm. We first used 5000 Metropolis
steps to bring the configuration to a local equilibrium, and then
continued with 3000 Wolff cluster formations before we started to collect
data. The data were averaged over 100 000 cluster formations, and after
every 1000 cluster moves ten
thermalizing Metropolis steps were completed.
\medskip

The flow of the fourth order cumulants is depicted in Fig. 4.
We can readily see that the value of $K$ at which the transition takes
place is $K_{I} \in [4.9,5.0]$.
A more detailed extraction of $K_{I}$ was done by studying the
ratio $U_{L^\prime_B}/U_{L_B}$ where $L^\prime_B \in [L_B,64]$ and
$L_B=2,4,6,8,12,16,20,24,32,48,64$.
The point where $U_{L^\prime_B}/U_{L_B} = 1$
was taken as the transition point. This extraction finally
gave the result

\begin{equation}
\label{eq:IsingMC}
K_{I} = 4.95 \pm 0.05.
\end{equation}

By flipping the sign of the four fold field,
we also confirmed numerically
that the Ising transition boundaries
are symmetrical with respect to the line $h_4=0$ (cf. Fig. 2(b)).
It is also possible to use a different
order parameter ($<\cos \phi_i \sin \phi_i>$) by rotating the spins
into the first quadrant. The results using
this order parameter agreed with the choice of $<\sin \phi_i>$.
We also used the order parameter  blocks to estimate the critical
exponent $\nu$  for the transition point, and obtained $\nu=1.1\pm0.1$
in very satisfactory agreement with the exact Ising result of
$\nu=1$.

\section{Monte Carlo Transfer Matrix Method}

Besides Monte Carlo simulations, another numerical
approach which has been quite
successful in the study of statistical mechanical
models is the transfer matrix method \cite{Nightingale90}. In this method,
the free energy of the model defined on an infinite strip can be obtained
directly
from the largest eigenvalue $\lambda_0$ of the transfer matrix as
$-\log \lambda_0$. These calculations can be done exactly for models with
discrete degrees of freedom when the transfer matrix is of low order. When
this technique is combined with finite size
scaling, one can obtain very accurate estimates of critical exponents and
other quantities. For models with higher
order transfer matrix or continuous degrees of freedom, as for the
case of the $XY$ model with symmetry breaking fields,
one has to resort to a Monte Carlo Transfer Matrix (MCTM) technique
\cite{NightingaleB88} in order to
estimate the largest eigenvalue of the transfer matrix. Some models
with continuous symmetry have
already been studied by this method, including frustrated $XY$
\cite{Granato92,GranatoN93} and coupled $XY$--Ising models
\cite{GranatoKLN91}.
\medskip

We proceed to summarize the method. Further details can found in
Refs. \cite{Nightingale90,NightingaleB88,GranatoN93,Nightingale88}.
The MCTM consists in a stochastic implementation of the well--known power
method to obtain the dominant eigenvalue of a matrix. First, helical
boundary conditions are implemeted in order to get a sparse transfer
matrix. At each step a new configuration is
obtained from the previous one, by adding a new spin $s^\prime_{L+1}$ and
relabeling the sites. The infinite strip can be constructed by repetition
of this identical elementary steps. This process
defines the transfer matrix to add a
single site $T(s^\prime,s)$, where $s={s_1,s_2,...,s_L}$ represent a
configuration of a column with $L$ spins. Then a sequency of random walkers
$R_i = {s_{1,i},s_{2,i},... s_{L,i}}$, $1 \le i \le r$, representing the
configurations of a column is introduced with corresponding weights
$w_i$. The number of walkers $r$ is maintained within a few percent of
a target value $r_0$ by adjusting the
weights properly. A matrix multiplication
can be regarded as a transition process from $s$ to $s^\prime$ with a
probability density $P(s^\prime,s)$ defined from the elements of the
transfer matrix as $T(s^\prime, s) = D(s) P(s^\prime,s)$
with a
normalization factor $D(s)$ independent of $s^\prime$. In each step
the weights are changed according to $w^\prime_i = D(s) w_i/c$
where
$c = \lambda_0 r / r_0$ is chosen to maintain $r$ close to $r_0$,
with $\lambda_0$ a running estimate of the eigenvalue. In this procedure
an MC step consists of a complete sweep over all random walkers.
After disregarding $t_0$ MC steps for equilibration,
an estimate of the largest eigenvalue can be obtained as

\begin{equation}
\lambda_0 = {{\displaystyle \sum_{t=t_0+1}^{T} \textstyle
c_{t+1} W_{t+1} }
\over { \displaystyle \sum_{t=t_0+1}^T \textstyle W_t }},
\end{equation}

where $W_t = \sum_i w_{i,t}$, with $w_{i,t}$ denoting the
configuration weights of a column at time $t$, and $T$ is the
total number of MC steps.

\medskip
For the calculations of the free energy of our model we
performed extensive runs using typically $r_0 = 20\  000$ random walkers
and $T=100 \ 000$ MC steps which corresponds to
$2 \times 10^9$ attempts per spin. We concentrated our attention
in two quantities, the interfacial free energy and the central charge.
The former can obtained from the free energy per site
$f = -\ln \lambda_0$, as calculated from the transfer matrix  on
an infinite strip of width $L$,
by a suitable choice of the boundary conditions.
If antiperiodic boundary conditions are used instead of periodic ones, an
interface is favored along the infinite strip and the associated
interfacial free energy $\Delta F(L)$ can be obtained from the free energy
difference between periodic and antiperiodic boundary conditions as

\begin{equation}
\Delta F(L) = L^2 \Delta f.
\end{equation}

The finite size scaling behavior of this quantity
should drastically change as a function of temperature depending on the
relevance of the symmetry breaking fields.
At low enough temperatures this
interface corresponds to a sharp boundary between two different phases
due to the presence of the symmetry breaking fields which are relevant at
these temperatures,
and should therefore increase as  $L^{d-1}$. On the other
hand, at higher temperatures when the symmetry breaking field is irrelevant,
the interface correspond to gradual change, in form of a twist
of the angle $\phi$, and $\Delta F(L)$ scales as $L^{d-2}$,
which is roughly a constant at $d=2$. In fact, in this regime
$\Delta F$ ($\approx const.$)
can be related to the helicity modulus (divided by $k_BT$)
$\gamma$ as

\begin{equation}
\gamma = 2 \Delta F/ \pi^2,
\end{equation}

for large enough $L$. The change in behavior
of the interfacial free energy between these regimes can be used to find
the multicritical point.

\medskip
Another important quantity which can be
inferred from the MCTM calculations,
is the central charge $c$, which classifies the possible conformally
invariant critical theories \cite{BloteCN86}. For example, $c=1/2$ for the
Ising model and $c=1$ along  the critical line of the $XY$ model.
This quantity can be related to the
amplitude of the singular part of the free energy per site
(at criticality) of the infinite strip by

\begin{equation}
f(K_c,L) = f_0 + {{\pi c}\over {6 L^2}}
\end{equation}

for sufficiently
large $L$, where $f_0$ denotes the regular contribution to the free energy.
Although $c$ is only defined at criticality, the value of $c$ extracted
from a $ f(K,L) \times {1/{L^2}}$  fitting of the free energy as
a function of system size, can be used to define an
effective size and coupling dependent central charge
$c(K,L)$ away from the critical point. According to
the Zamolodchikov $c$-theorem \cite{Zamolodchikov}, this quantity
should have a well defined behavior near the critical point
and reach a constant value, equal to the central charge,
at the fixed point. As a consequence $c(K,L)$ should have a maximum near
an  unstable fixed point
and converge to $c=0$ in the completely disordered
or ordered phases. This property is particularly useful in
locating the multicritical point and the Ising transition.
\medskip

To facilitate direct comparison with the MC results of Sec. 3,
we set $h_4=0$ and $h_8=-0.15$.
The results of the MCTM calculations for the central charge
in this non--competing case (cf. Fig. 2(a)) are shown in Fig. 5.
The value of $c$ was estimated by fitting the free energy
per site to Eq. (24), using sizes $L=5 -- 11$.
First, the abrupt onset of $c$ near $K\approx 1$ agrees well
with the known result for the $XY$ transition (where $h_8$ is
irrelevant).  The multicritical point, on the other hand
is estimated to be at the point where the central charge begins to
decrease from the value of one (the $XY$ phase value) to zero
at $K_m = 2.6\pm 0.4$.
This is in fair agreement with the result of Eq. (16)
as obtained from the
histogram method; the discrepancy can be
attributed to severe finite size effects for the relatively
small strip widths studied.

\medskip
In Fig. 6, we show the results
for the interfacial free energy for different strip widths $L$ as a function
of temperature for the same non--competing case. One clearly sees a
size dependence at large values of $K$ indicating
the relevance of the symmetry breaking field $h_8$. At temperatures above
$K_m^{-1} \approx 1/2.6$ the size
dependence is practically absent indicating
the spin wave regime of the critical line of the $XY$ model. In this regime
the helicity modulus is equal to the
renormalized coupling constant $K^*$
and should be equal to $K^* = 8/\pi$ at $K_{m}$.
As indicated in Fig. 6, the helicity modulus
$\gamma = 2 \Delta F/ \pi^2$ is in
good agreement with the expected result. Moreover, the termination
point of the lines at $\Delta F=0$  is in good agreement with
the KT transition point.
\medskip

Next, we use the MCTM method to study the Ising transition in
the case of competing $h_4$ and $h_8$ fields. As in Sec. 4,
we set $h_4=0.06$ and $h_8=0.15$. The results for $c$ {\em vs.} $K$
are shown in Fig. 7.
The effective central charge has two peaks as a function
of temperature. At high temperatures the peak value is very close to
$c=1$, which is consistent with a transition in the universality
class of the $XY$ model with a four fold field.
The other peak at low temperatures, is close to $c=1/2$ which
is consistent with an Ising transition.
The Ising transition can be estimated simply from the
peak value of $c$ yielding $K_I = 4.6 \pm 0.4$.
These results are again in reasonable agreement with
the MCRG method, Eq. (20).

\section{Summary and Discussion}

In this work, we have analysed in detail the properties of the
two dimensional $XY$
model in the presence of four fold and eight fold symmetry breaking fields
$h_4$ and $h_8$, respectively.
First we have applied mean  field and
renormalization group arguments to
predict that when the $h_4$ field changes from positive to negative values,
there is a phase transition between a phase with the
order parameter pointing along the $\phi=\pi/4$ direction,
and another phase with the corresponding order
parameter pointing along the $\phi=0$ direction. This phase transition is
continuous at high temperatures where the eight
fold field is irrelevant. In
this case, right at the phase boundary all the symmetry breaking fields are
absent and the system is in the ordered phase
of a pure $XY$ model with only
algebraic long range order. At lower temperatures when the eight fold field
is relevant, the nature of the transition now depends crucially on whether
$h_8$ and $h_4$ are compatible or
competing with each other. In the former case,
the transition becomes first order and there is a multicritical temperature
separating the high temperature and low temperature phase boundary. In the
competing case the first order
phase transition splits into two transitions both of
which are in the Ising universality class. The order parameter rotates
continuously from the $\phi = 0$ direction towards the
$\phi = \pi/4$ direction in between the
two Ising phase boundaries.
We note that similar effects can result from the
competition of six fold and twelve fold
fields as discussed by Selinger and
Nelson \cite{Sel88} in their study of liquid crystals.
\medskip

We have then performed detailed numerical analysis of our model
Hamiltonian through Monte Carlo simulations. Because of the large finite
size effects, the finite size simulation data has to be extrapolated to
infinite size through finite size scaling concepts and numerical
renormalization group analysis. We have
first confirmed the qualitative nature of
the phase boundary as discussed above. We have then employed
the recently developed
histogram techniques to locate the multicritical temperature. Finally, in the
case of the competing fields, we have used Binder's
block renomalization technique
to identify the Ising like transitions. These results
have further been corroborated by Monte Carlo Transfer Matrix
techniques. All our numerical results
are in good agreement with theoretical arguments.
\medskip

Our model was originally motivated by the study of the $H/W(100)$
chemisorption system.
We have shown previously that the critical properties of
this system in the ordered $c(2\times 2)$
phase can be desribed by an $XY$ model with
four and eight fold symmetry breaking fields. The effect of the
adsorbed hydrogen in the low coverage limit is to change the effective
four fold field. Thus increasing the hydrogen coverage in this system is
tantamount to changing the four fold
field from negative to positive values in
the model Hamiltonian studied in this paper. In addition, our
previous work showed \cite{Kan93b}
that the adsorbed hydrogen also generates
an eight fold field which is compatible with the four fold
field. Experimental evidence from an
infrared spectroscopy study \cite{Arr86}
for this system supports the scenario of a
continuous transition
from $<11>$ (corresponding to negative $h_4$) to $<10>$ (positive $h_4$)
phase at room temperature when the hydrogen coverage is
increased. On the other hand, the Low Energy Electron Diffraction
study of Griffiths {\it et al.} \cite{Gri81} showed
indications of coexistence between
the two phases for coverages in the range from about
0.05 to 0.16 which indicates a first order transition. This is exactly the
behavior of our model Hamiltonian studied
in this paper when the fields $h_4$
and $h_8$ are compatible.  Thus the qualitative agreement between the
experimental data and the theory presented here is gratifying. In view of
the very rich behaviour of the phase diagram, especially near the
multicritical point, more experimental
studies of the switching transition
in this system and comparison with the theoretical
predictions here would be most fruitful.

\section { Acknowledgments }

This work has been supported by the Conselho
Nacional de Desenvolvimento Cient\'ifico e Tecnol\'ogico (CNPq) --
NSF International Collaboration program.
T.A--N. and K.K. have been partly supported by the Academy of Finland,
S.C.Y. by an ONR grant, and J.M.K. by an NSF grant.
E.G. acknowledges the support from Funda\c c\~ao de Amparo \`a Pesquisa
do Estado de S\~ao Paulo (FAPESP, No. 92/0963-5) and
CNPq.
The Center for Scientific Computing and Tampere University
of Technology are acknowledged for the computational resources.

\clearpage
\Large
{\bf Figure Captions}
\normalsize
\vskip1.0cm
\baselineskip 20pt

Fig. 1.(a) A schematic figure of the anisotropy potential $V(\phi)$
           of Eq. (2), for the case of non--competing fields with
           $h_8=-1$. The curves from bottom to the top are for
           $h_4=-6$, $-4$, $-2$,  0, 2, 4, and 6, respectively.
           The transition at $h_4=0$ is abrupt.
       (b) The anisotropy potential for competing fields  with
           $h_8=+1$. The curves are for the same values of
           $h_4$ as in (a). The transition around $h_4=0$ is
           continuous.
\medskip

Fig. 2. (a) A schematic phase diagram for the Hamiltonian Eq. (1)
            on the $h_4-K^{-1}$ plane, with a fixed non--competing
            $h_8 <0$. For finite values of $h_4$, the two transition
            lines  beyond the Kosterlitz--Thouless
            transition point $K^{-1}_{KT}<\pi/2$
            belong to the universality
            class of the $XY$ model with a four fold symmetry breaking
            field. Between $K^{-1}_{KT}$ and the multicritical point
            $K^{-1}_m$, there is a line of pure $XY$ transitions.
            Below $K^{-1}_m$, the eight fold field is relevant and
            there is a line of first order transitions. The lower
            bound for $K^{-1}_m$ is $\pi/8$ (cf. Eq. (9)).
        (b) The counterpart of the phase diagram of (a) for the case
            of a fixed, competing $h_8>0$. Below $K^{-1}_m$, where
            $h_8$ is relevant, the first order transitions open up
            into two lines of continuous Ising transitions, which
            terminate at $\pm 4\vert h_8 \vert$.
\medskip

Fig. 3. (a) Logarithm of the histogram of the local order parameter
	    $\Phi = (1/L^2) \sum_i \cos4\phi_i$ (cf. Eq. (\ref{eq:defF})),
            at $K=2.1$ for system
	    sizes $L=8,12,16,48,64$, with $h_4=0.06$ and $h_8=-0.15$.
            $N$ is the number of samples in each bin.
	(b) Logarithm of the histogram of the local order parameter
	    $\Phi = (1/L) \sum_i \cos4\phi_i$
            at $K=2.4$ for system
	    sizes $L=8,12,16,24,48,64$, with $h_4=0.06$ and $h_8=-0.15$.
        (c) The free energy barriers $\Delta F(L)$ (cf. Eq. (15))
            extracted from polynomial fits to the histograms of
            the order parameter, plotted against $L$. Here
            $h_4=0$ and  $h_8=-0.15$, corresponding to Fig.
            2(a). Above
            $K_m\approx 2.2$, the barriers clearly start increasing
            indicating the onset of a first order transition.
            See text for details.
\medskip

Fig. 4. Typical MCRG flows of the fourth order
        cumulant {\em vs.} $1/L$ for the
        competing case, with $h_4=0.06$ and $h_8=0.15$. When
        $K$ is varied, the Ising transition occurs at around
        $K_I \approx 4.9-5.0$ (cf. Fig. 2(b)). See text  for
        details.
\medskip

Fig. 5. MCTM  results for the effective central charge along the line
        $h_4=0$, with $h_8=-0.15$ (the non--competing case of
        Fig. 2(a)). The first abrupt
        onset to $c=1$ corresponds to the KT transition, and the
        onset of decline of $c$ at $K_m \approx 2.6$  indicates entry
        to the line of first order phase transitions
        just below $K^{-1}_m$ in Fig. 2(a).
\medskip

Fig. 6. The interfacial free energy from the MCTM method for different
        strip sizes as a function of $K$, for the case $h_4=0$ and
        $h_8=-0.15$ as in Fig. 5.
        Below $K_m \approx 2.6$, the lack of size dependence
        of the energy indicates the onset the spin wave regime
        corresponding to a line of $XY$ transitions.
        The corresponding
        value of $K^*$ is in good agreement with the theoretical
        prediction $8/\pi$. See text for details.
\medskip

Fig. 7. MCTM results for the effective central charge with $h_4=0.06$,
        $h_8=0.15$, corresponding to the competing case of Fig. 2(b).
        The first peak at $c=1$ corresponds to the KT transition
        point, while the second peak at $K_I \approx 4.6$ verifies
        the expected Ising transition, with $c=1/2$.

\end{document}